\documentstyle[preprint,aps]{revtex}
\input epsf

\def\la{\left\langle}
\def\ra{\right\rangle}

\begin{document}
\draft

\title{Dynamical Supersymmetric  Inflation}
\author{William H.\ Kinney\thanks{Electronic mail: {\tt kinneyw@fnal.gov}} and Antonio Riotto\thanks{EPPARC Advanced Fellow,  Oxford Univ. from Sept.~1997.
{}From 1 Dec.~1997 on leave of absence at CERN, Theory Division, CH--1211
Geneva 23, Switzerland. Electronic mail: {\tt riotto@fnas01.fnal.gov}.}}
\address{\em NASA/Fermilab Astrophysics Center \\
Fermi National Accelerator Laboratory, Batavia, IL~~60510}
\author{FERMILAB--PUB--97/090--A}
\author{hep-ph/9704388}
\date{April 23, 1997}
\maketitle

\begin{abstract}
We  propose a new class   of inflationary models  in which the scalar field potential governing inflation is generated by the same non-perturbative gauge dynamics that may lead to supersymmetry breaking. Such models satisfy constraints from cosmic microwave background measurements for natural values of the fundamental parameters in the theory. In addition, they have two particularly interesting characteristics: a ``blue'' spectrum of scalar perturbations, and an upper bound on the total amount of inflation possible.
\end{abstract}


\section{Introduction}

The existence of an inflationary stage during the evolution of the early
Universe is usually invoked to solve the flatness and the horizon problems of
the standard big  bang cosmology \cite{guth81}. During inflation the energy
density is dominated by vacuum energy and comoving scales undergo quasi-exponential growth. As a result, any undesirable topological defects left as remnants after
some Grand Unified phase transition, such as monopoles, are diluted.
Typically, the vacuum energy driving inflation is generated by
a scalar field $\phi$ (the ``inflaton'') displaced from the minimum of a 
potential $V\left(\phi\right)$. Quantum fluctuations of the inflaton field imprint a
nearly scale invariant spectrum  of fluctuations on the background space-time
metric. These fluctuations may be responsible for the generation of structure
formation. However, the level of density and temperature fluctuations observed
in the present Universe, $\delta\rho/\rho\sim 10^{-5}$, require the inflaton
potential to be extremely flat.
 For instance, in the chaotic inflationary scenario \cite{chaotic} where the
inflaton potential is $V=\lambda\phi^4$ and the scalar field sits initially at
scales of order of the Planck scale, the dimensionless self-coupling $\lambda$
must be of order of $10^{-13}$ to be consistent with observations.
 The inflaton field must be coupled to other fields in order to ensure the
conversion of the vacuum energy into radiation at the end of inflation, but
these couplings must be very small, otherwise loop corrections to the inflaton
potential spoil its flatness. While the necessity of introducing very small
parameters to ensure the extreme flatness of the inflaton potential seems very
unnatural and fine-tuned in  most non-supersymmetric theories, this technical
naturalness may be achieved in supersymmetric models \cite{ellis82} because the
nonrenormalization theorem guarantees that the superpotential is not
renormalized to all orders of perturbation theory \cite{grisaru79}. However,
even though the form of the potential may be stable when the inflaton  couples
to other fields, initial small couplings $\lambda$ must still be put by hand, 
an aesthetically unpleasant option. A natural way to solve this problem is to
interpret small couplings as a ratio of different mass scales. This is the
underlying idea of hybrid inflation \cite{linde91,linde94} where two (or more)
interacting fields are associated with different scales and inflation
ends  by a rapid rolling of a second field $\sigma$, initially sitting at the origin
and  triggered by the slow rolling of the field $\phi$. During the inflationary
stage the inflaton potential is typically  given by  $V=V_0+
\frac{m^2}{2}\phi^2$ and the smallness of the density perturbations may be
easily explained by the hierarchy \hbox{$m < V_0^{1/4}\ll M_{Pl}$}. Successful
supersymmetric versions of hybrid inflation have been constructed in the
framework of global supersymmetry \cite{shafi} and, more recently, in
supergravity theories \cite{sugra}. Supersymmetry can therefore play a fundamental role  during inflation \cite{lyth96}.

In  our discussion of inflation below we will  concern ourselves with a potential of the form:
\begin{equation}
\label{pot}
V(\phi)=\frac{\Lambda_3^{p+4}}{\phi^p},
\end{equation}
where the index $p$ and the scale $\Lambda_3$ depend upon the underlying gauge
group. 
Such a potential might at first seem peculiar since it involves
a field (or more than one field) appearing in the denominator. However, such 
potentials are known to arise generically in supersymmetric theories \footnote{ Strictly speaking we are referring here to the 
superpotential in a supersymmetric theory. When the superpotential arises due to instanton effects, and
 when the gauge groups involved are completely broken at a high
enough scale, one can argue that the resulting potential  has a similar form.}. 
This fact has been known for some time \cite{ads},
and has gained increasing attention of late after the recent work on understanding the non-perturbative behavior of supersymmetric  gauge theories \cite{review}. The unconventional term 
involving fields in the denominator typically arises due to 
non-perturbative effects that lift various flat directions in 
supersymmetric theories. Such non-perturbative effects might have to do with 
instanton effects or, for example, with gaugino condensation associated with 
some unbroken non-Abelian gauge group. It is also worth mentioning that regardless of exactly how the potential in Eq. (1) arises, in our discussion of 
inflation we will be working in the 
region where the field(s) involved have large enough vacuum expectation values
that their presence in the denominator  does not cause any singular behavior. 
In this regime the theory is weakly coupled and corrections to the canonical Kahler potential are negligible.

We will not be concerned, in this paper,  with the the detailed dynamics 
which gives rise to the  potential in Eq. (1). This is in large part because
our conclusions are quite insensitive to most details of the 
potential and follow mainly from the {\em general} form of Eq. (1). 
It is of course interesting to ask whether the inflationary scenario 
envisaged in this paper can be actually realized in a reasonably simple 
supersymmetric model.  It has been suggested, 
for example in models of gauge mediated supersymmetry breaking,  that 
non-perturbative effects, driven by instanton effects or gaugino condensation, 
might be responsible for the dynamical supersymmetry breaking (DSB) when  
the classical moduli space is lifted in 
the 
potential and scalar fields are driven to large expectation values. Additional
tree level interactions in the potential may raise the potential at large expectation
values leading to a stable ground state. If some $F$-terms (and the potential) do
not vanish in such  a  ground state, supersymmetry is spontaneously broken. 
Recent developments have also shown that many supersymmetric theories may have
other types of non-perturbative dynamics which lead to degenerate quantum moduli
spaces of vacuum instead of dynamically generated superpotentials.

This makes it  tempting to speculate (although it is not necessary for the 
discussion of inflation proposed in this paper) that the potential above, responsible for inflation,  could also arise from the 
same underlying dynamics responsible for supersymmetry breaking.
We will not pursue these ideas any further here and leave them for future study. We only note in passing   that 
the term in Eq. (\ref{pot})
is not usually the only one present in the potential. Other terms may be
present and  lift up the flat direction.  If  these terms are small enough,  the vacuum
expectation value  of the field is  finite, but nevertheless  all the
interesting dynamics of  inflation  happens when the term (\ref{pot})
dominates. This comes out naturally  if these other terms   are
nonrenormalizable, which is  plausible in many DSB models.
Moreover, following the philosophy of such models, it is likely that
nonrenormalizable terms  are suppressed by powers of $M_{Pl}$,  the only
explicit scale  allowed in the theory.

The last key ingredient of our proposal is that during inflation,  the energy
density  is dominated by a a nonzero constant vacuum energy density $V_0$ (this
always happens if the value of the inflaton field is {\it large} enough). In
the absence of any other physics, the inflaton field rolls down to its VEV and
inflation never ends. However, the inflaton may be coupled to  some fields in
some other sector of the theory in such  away that, when $\phi$ gets larger
than some critical value $\phi_c$, $V_0$ drops to zero. This is the usual way
inflation is terminated in hybrid inflation. What is remarkable is that this
vacuum energy density $V_0$ may be identified with some scale $\Lambda_2^4$
which appears in models where  quantum deformation of a classical moduli space
plays a fundamental role. A typical example is provided by the chiral
$SU(2)\otimes  SU(3)$ model in the limit in which the scale associated to
$SU(2)$ is much higher than the one associated to $SU(3)$. Here we are assuming
that the scale $V_0$ is only present during the inflationary stage and relaxes
to zero to opportunely end inflation.

In this paper we propose a new class   of inflationary
models  inspired by the typical structure of the potential generated by some
non-perturbative gauge dynamics.   We will denote this class of models by
dynamical supersymmetric inflation (DSI).
We will show that a successful inflationary scenario  may be constructed and
that the generation of density perturbations may be accounted for in the limit
$\Lambda_2\gg \Lambda_3$. More interestingly, a blue spectrum of density
perturbations is predicted.

The paper is organized as follows:  Section (\ref{secinflationreview}) is a
short review of inflationary cosmology from scalar field theories. Section
(\ref{secinflationcalc}) contains the details of limiting the parameters of the
model from observations of CMB fluctuations. Finally, Section
(\ref{secconclusions}) contains a summary and conclusions.

\section{Inflation in scalar field theories}
\label{secinflationreview}

In this section, we quickly review scalar field models of inflationary
cosmology, and explain how we relate model parameters to observable quantities. (For more detailed reviews, see Refs. \cite{lidsey95,lyth96}.)
If the stress-energy of the universe is dominated by a scalar field with potential $V\left(\phi\right)$, the Einstein Field equations $G_{\mu\nu} = \left(8 \pi / M_{Pl}^2\right) T_{\mu\nu}$ for the evolution of the background metric reduce to
\begin{equation}
H^2 \equiv \left({\dot a \over a}\right)^2 = {8 \pi \over 3 M_{Pl}^2} \left[{1 \over 2} \dot\phi^2 +
V\left(\phi\right)\right].
\label{eqbackground}
\end{equation}
Here $a\left(t\right)$ is the scale factor, and $M_{Pl} = G^{-1/2} \simeq 10^{19}\ {\rm GeV}$ is the Planck mass. {\em Inflation} is defined to be a period of accelerated expansion, $\ddot a > 0$.
The evolution of the scale factor can be written as $a \propto e^N$, where the number of e-folds $N$ is defined in terms of the Hubble parameter $H$ as
\begin{equation}
N \equiv \int{H\,dt}.
\end{equation}
During inflation $H$, and therefore the horizon size $d_H \simeq H^{-1}$, is
nearly constant, and the expansion of the universe is quasi-exponential. This
results in the curious behavior that the coordinate system is expanding faster
than the light traveling in it, and comoving length scales rapidly increase in
size relative to the horizon distance. Regions initially in causal contact are
``redshifted'' to large, non-causal scales, explaining the observed isotropy of
the cosmic microwave background (CMB) on large angular scales. This is also
important for the generation of metric fluctuations in inflation, discussed
below. Finally, a universe which starts out with a nonzero curvature evolves
rapidly during inflation toward zero curvature and a flat Robertson-Walker
metric.

Stress-energy conservation gives the equation of motion of the scalar field
\begin{equation}
\ddot\phi + 3 H \dot\phi + V'\left(\phi\right) = 0.
\label{eqfieldeom}
\end{equation}
The {\em slow-roll} approximation\cite{linde82,albrecht82} is the assumption
that the evolution of the field is dominated by drag from the cosmological
expansion, so that $\ddot\phi \simeq 0$ and
\begin{equation}
\dot \phi \simeq -{V' \over 3 H}.
\end{equation}
The equation of state of the scalar field is then dominated by the potential,
so that $p \simeq -\rho$, and the expansion rate is approximately
\begin{equation}
H \simeq \sqrt{{8 \pi \over 3 M_{Pl}^2} V\left(\phi\right)}.
\label{eqhslowroll}
\end{equation}
The slow-roll approximation is consistent if both the slope and curvature of
the potential are small. This condition is conventionally
expressed in terms of the ``slow-roll parameters'' $\epsilon$ and $\eta$, where
\begin{equation}
\epsilon \equiv {M_{Pl}^2 \over 4 \pi} \left({H'\left(\phi\right) \over
H\left(\phi\right)}\right) \simeq {M_{Pl}^2 \over 16 \pi}
\left({V'\left(\phi\right) \over V\left(\phi\right)}\right)^2,
\end{equation}
and
\begin{equation}
\eta\left(\phi\right) \equiv {M_{Pl}^2 \over 4 \pi} \left({H''\left(\phi\right)
\over H\left(\phi\right)}\right) \simeq {M_{Pl}^2 \over 8 \pi}
\left[{V''\left(\phi\right) \over V\left(\phi\right)} - {1 \over 2}
\left({V'\left(\phi\right) \over V\left(\phi\right)}\right)^2\right].
\end{equation}
Slow-roll is then a consistent approximation for $\epsilon,\ \eta \ll 1$. The
parameter $\epsilon$ can be shown to directly parameterize the equation
of state of the scalar field, $p = -\rho \left(1 - 2/3 \epsilon\right)$, so
that the condition for inflation $\ddot a > 0$ is
{\em exactly} equivalent to $\epsilon < 1$. The number of e-folds $N$ of
inflation as the field evolves from $\phi_i$ to $\phi_f$ can be expressed
in terms of  $\epsilon$ as
\begin{equation}
N = {2 \sqrt{\pi} \over M_{Pl}} \int_{\phi_i}^{\phi_f}{d\,\phi \over
\sqrt{\epsilon\left(\phi\right)}}.
\end{equation}
To match the observed degree of flatness and homogeneity in the universe, we
require many e-folds of inflation, typically $N \simeq 50$. (This figure varies
somewhat with the details of the model.)

Inflation not only explains the high degree of large-scale homogeneity in
the universe, but also provides a mechanism for explaining the observed {\em
inhomogeneity} as well. During inflation, quantum fluctuations on small scales
are quickly redshifted to scales much larger than the horizon size, where they
are ``frozen'' as perturbations in the background
metric\cite{hawking82,starobinsky82,guth82,bardeen83}. Metric perturbations at
the surface of last scattering are observable as temperature anisotropy in the
CMB, which was first detected by the Cosmic Background Explorer (COBE)
satellite. The metric perturbations created during inflation are of two types:
scalar, or {\it curvature} perturbations, which couple to the stress-energy of
matter in the universe and form the ``seeds'' for structure formation, and
tensor, or gravitational wave perturbations, which do not couple to matter.
Both scalar and tensor perturbations contribute to CMB anisotropy. Scalar
fluctuations can also be interpreted as fluctuations in the density of the matter in the universe. Scalar fluctuations can be
quantitatively characterized by perturbations $P_{\cal R}$ in the intrinsic
curvature\cite{mukhanov85,mukhanov88,mukhanov92,stewart93}
\begin{equation}
P_{\cal R}^{1/2}\left(k\right) = {1 \over \sqrt{\pi}} {H \over M_{Pl}
\sqrt{\epsilon}}\Biggr|_{k^{-1} = d_H}.
\end{equation}
The fluctuation power is in general a function of wavenumber $k$, and is
evaluated when a given mode crosses outside the horizon during inflation,
$k^{-1} = d_H$. Outside the horizon, modes do not evolve, so the amplitude of
the mode when it crosses back {\em inside} the horizon during a later radiation
or matter dominated epoch is just its value when it left the horizon during
inflation. The {\em spectral index} $n_{\cal R}$ is defined by assuming an
approximately power-law form for $P_{\cal R}$ with
\begin{equation}
n_{\cal R} - 1 \equiv {d\ln\left(P_{\cal R}\right) \over d\ln\left(k\right)},
\end{equation}
so that a scale-invariant spectrum, in which modes have constant amplitude at
horizon crossing, is characterized by $n_{\cal R} = 1$. Instead of specifying
the fluctuation amplitude directly as a function of $k$, it is often convenient
to specify it as a function of the number of e-folds $N$ before the end of
inflation at which a mode crossed outside the horizon. Scales of interest for
measurements of CMB anisotropy crossed outside the horizon at $N \simeq
50$, so that $P_{\cal R}$ is conventionally evaluated at $P_{\cal R}\left({N =
50}\right)$. Similarly, the power spectrum of tensor fluctuation modes is given
by
\begin{equation}
P_{T}^{1/2}\left(k_N\right) = {4 \over \sqrt{\pi}} {H \over M_{Pl}}\Biggr|_{N =
50}.
\end{equation}
The ratio of tensor to scalar modes is 
$\left({P_{T} / P_{\cal R}}\right) =  16 \epsilon$, 
so that tensor modes are negligible for $\epsilon \ll 1$. If the contribution
of tensor modes to the CMB anisotropy can be neglected, normalization to the
COBE four-year data gives\cite{bunn96,lyth96} $P_{\cal R}^{1/2} = 5 \times
10^{-5}$.
Calculating the CMB fluctuations from a particular inflationary model reduces
to the following basic steps: (1) from the potential, calculate $\epsilon$ and
$\eta$. (2) From $\epsilon$, calculate $N$ as a function of the field $\phi$.
(3) Invert $N\left(\phi\right)$ to find $\phi_{N = 50}$. (4) Calculate $P_{\cal
R}$, $n_{\cal R}$, and $P_T$ as functions of $\phi$, and evaluate at $\phi_{N =
50}$ to determine the values of the observables at scales of current
astrophysical interest.

\section{Inflation from non-perturbative gauge dynamics in supersymmetry}
\label{secinflationcalc}

We take the potential to be  described by a single degree of freedom $\phi$, of
the general form
\begin{equation}
V\left(\phi\right) = V_0 + {\Lambda_3^{p + 4} \over \phi^p} + {\phi^{q + 4}
\over M_{Pl}^q}.
\end{equation}
As we noted in the introduction, the presence of the  nonrenormalizable
term is not strictly necessary. We include it here for generality, and show that its presence, subject to certain consistency constraints, does not significantly affect our conclusions.

The minimum of the potential is at $V'\left(\la\phi\ra\right) = 0$, where the
vacuum expectation value (VEV) $\la\phi\ra$ is given by
\begin{equation}
\la\phi\ra = \left[\left({p \over q + 4}\right) \Lambda_3^{p + 4}
M_{Pl}^q\right]^{1/(p + q + 4)}.
\end{equation}
Note that, in general, the potential does not vanish when the field is at the
VEV. This means that if the Universe ever becomes vacuum dominated, it {\em
stays} vacuum dominated, and inflation will continue indefinitely unless other
physics is brought into play. Here we will simply assume that some other sector
of the theory ends inflation when $\phi$ passes through a critical value
$\phi_c$. For $\phi_c$ near the VEV $\la\phi\ra$, the potential can be expanded
as
\begin{equation}
V\left(\phi\right) = V\left(\la\phi\ra\right) + {1 \over 2}
V''\left(\la\phi\ra\right) \left(\phi - \la\phi\ra\right)^2 + \cdots.
\end{equation}
This is just the case of standard hybrid inflation\cite{linde91,linde94}, which has been studied
extensively in the literature. Here we study the
limit $\phi \ll \la\phi\ra$, for which such an expansion is not possible. In
this limit, the $\phi^{-p}$ term dominates the dynamics,
\begin{eqnarray}
V\left(\phi\right) &&\simeq  V_0 + {\Lambda_3^{p + 4} \over \phi^p},\qquad \phi
\ll \la\phi\ra\cr
&&= V_0 \left[1 + \alpha \left(M_{Pl} \over \phi\right)^{p}\right],
\end{eqnarray}
where
\begin{equation}
\alpha \equiv {\Lambda_3^{p + 4} \over M_{Pl}^p V_0}.
\end{equation}
We assume that the constant $V_0$ dominates the potential, or $\alpha \ll
\left(\phi / M_{Pl}\right)^p$. In this limit, the first slow-roll parameter is
\begin{equation}
\epsilon\left(\phi\right) = {M_{Pl}^2 \over 16 \pi} \left({V'\left(\phi\right)
\over V\left(\phi\right)}\right)^2 = \left({\phi_0 \over \phi}\right)^{2
\left(p + 1\right)},\label{eqepsilon}
\end{equation}
where
\begin{equation}
\left({\phi_0 \over M_{Pl}}\right) = \left({p \over 4 \sqrt{\pi}}
\alpha\right)^{1/\left(p + 1\right)}.
\end{equation}
The second slow-roll parameter $\eta$ is
\begin{eqnarray}
\eta\left(\phi\right) &&= \epsilon - {M_{Pl} \over 4 \sqrt{\pi}} {\epsilon'
\over \sqrt{\epsilon}}\cr
&&= \left({\phi_0 \over \phi}\right)^{2 \left(p + 1\right)} + \left({p + 1
\over 2 \sqrt{\pi}}\right) \left({\phi_0 \over \phi}\right)^{p + 1}
\left({M_{Pl} \over \phi}\right).\label{eqeta}
\end{eqnarray}
Note that for $\phi \simeq \phi_0 \ll M_{Pl}$, the parameter $\eta$ becomes
large, indicating a breakdown of the slow-roll approximation. In particular, it
is inconsistent to say that inflation {\em begins} at $\phi = \phi_0$, when
$\epsilon\left(\phi\right)$ in Eq. (\ref{eqepsilon}) is equal to unity, since
that expression depends on the assumption of slow-roll. However, for $\phi \gg
\phi_0$, both $\epsilon$ and $\eta$ are small and slow-roll is a consistent
approximation. In the region $\phi_0 \ll \phi \ll M_{Pl}$, the second term in
(\ref{eqeta}) dominates, which is equivalent to $\eta \gg \epsilon$, and $\eta$
can be written in the useful forms
\begin{eqnarray}
\eta\left(\phi\right) &&\simeq {p + 1 \over 2 \sqrt{\pi}}
\sqrt{\epsilon\left(\phi\right)} \left({M_{Pl} \over \phi}\right)\cr
&&= {p \left(p + 1\right) \over 8 \pi} \alpha \left(M_{Pl} \over \phi\right)^{p
+ 2}.\label{eqformsofeta}
\end{eqnarray}
The number of e-folds $N$ is given by
\begin{eqnarray}
N = {2 \sqrt{\pi}\over M_{Pl}} \int_{\phi}^{\phi_c}{d\,\phi' \over
\sqrt{\epsilon\left(\phi'\right)}} &&= \left({p + 1 \over p + 2}\right) {1
\over \eta - \epsilon}\Bigg|_\phi^{\phi_c}\cr
&&\simeq \left({p + 1 \over p + 2}\right) \left({1 \over
\eta\left(\phi_c\right)} - {1 \over \eta\left(\phi\right)}\right),\quad
\epsilon \ll \eta,
\end{eqnarray}
where $\phi_c$ is the critical value at which inflation ends. The value of
$\phi_c$ is in general determined by a coupling of the field $\phi$ to some
other sector of the theory which we have here left unspecified. Accordingly, we
will treat $\phi_c$ as simply a free parameter. Noting from Eq.\
(\ref{eqformsofeta}) that $\eta \propto \phi^{-\left(p + 2\right)}$, for $\phi
\ll \phi_c$ the number of e-folds $N$ approaches a constant, which we call
$N_{\rm tot}$,
\begin{equation}
N_{{\rm tot}} \equiv  \left({p + 1 \over p + 2}\right) {1 \over
\eta\left(\phi_c\right)} = {8 \pi \over p \left(p + 2\right)} \alpha^{-1}
\left({\phi_c \over M_{Pl}}\right)^{p + 2}.
\end{equation}
This is quite an unusual feature. Most models of inflation have no intrinsic upper limit on the total amount of expansion that takes place during the inflationary phase, although only the last $50$ or $60$ e-folds are of direct observational significance. Here the total amount of inflation is bounded from above, although that upper bound can in principle be very large.
Defining $\phi_N$ to be the field value $N$ e-folds before the end of
inflation, we can then write $\eta\left(\phi_N\right)$ in terms of $N$ and
$N_{\rm tot}$ as
\begin{equation}
\eta\left(\phi_N\right) = \left({p + 1 \over p + 2}\right) {1 \over N_{\rm tot}
 - N },
\end{equation}
so that $\eta$ approaches a constant value for $N \ll N_{\rm tot}$. The
magnitude of scalar metric perturbations is given by the curvature power
spectrum $P_{\cal R}$,
\begin{eqnarray}
P_{\cal R}^{1/2} &&\equiv {1 \over  \sqrt{\pi}} {H\left(\phi_{50}\right) \over
M_{Pl} \sqrt{\epsilon\left(\phi_{50}\right)}}\cr
&&= {\left(p + 1\right) \over  \pi} \sqrt{2 \pi \over 3} \left(V_0^{1/2} \over
M_{Pl} \phi_{50}\right) {1 \over \eta\left(\phi_{50}\right)}\cr
&&= {\left(p + 2\right) \over  \pi} \sqrt{2 \pi \over 3} \left({V_0^{1/2} \over
M_{Pl} \phi_c}\right) N_{\rm tot} \left(1 - {50 \over N_{\rm
tot}}\right)^{\left(p + 1\right) / \left(p + 2\right)},
\end{eqnarray}
The COBE normalization is \cite{bunn96,lyth96} $P_{\cal R}^{1/2} = 5 \times
10^{-5}$, with spectral index
\begin{eqnarray}
n_{\cal R} - 1 &&\equiv {d\,\log\left(P_{\cal R}\right) \over
d\,\log\left(k\right)} =  - 4 \epsilon + 2 \eta\cr
&&\simeq \left({p + 1 \over p + 2}\right) {2 \over N_{\rm tot} \left(1 - 50 /
N_{\rm tot}\right)}.
\end{eqnarray}
As announced, the spectrum turns out to be blue, but  for $N_{\rm tot} \gg 50$
the spectrum approaches scale-invariance, $n_{\cal R} \simeq 1$. If we take the
example case of $p = 2$ and $\phi_c \sim V_0^{1/4}$, the COBE constraint on
$P_{\cal R}$ is met for $V_0^{1/4} \simeq 10^{10}\ {\rm GeV}$ and $\Lambda_3
\simeq 10^{6}\ {\rm GeV}$, very natural values for the fundamental scales in
the theory. Since $V_0^{1/4} \ll M_{Pl}$, tensor modes produced during inflation are of negligible amplitude, a typical feature of hybrid inflation models. 

In this class of models, a nearly scale-invariant spectrum $n_{\cal R} \simeq
1$ is the most natural case. Significantly blue spectra, $n_{\cal R} > 1$, are,
however, not excluded and can occur for properly tuned values of the
fundamental scales in the theory. The condition for significant deviation from
a scale-invariant spectrum is that $N_{\rm tot}$ not be much greater than $50$.
We can write $N_{\rm tot}$ as a function of the scalar spectral index
\begin{equation}
N_{\rm tot} = 50 + \left({p + 1 \over p + 2}\right) {2 \over n_{\cal R} - 1},
\end{equation}
so, again taking $p = 2$, a spectral index of $n_{\cal R} > 1.1$ requires
$N_{\rm tot} < 65$. Such a small amount of inflation could have observationally important consequences\cite{berera97}.

We have two observationally determined constraints, $P_{\cal R}^{1/2} = 5 \times 10^{-5}$ and $n_{\cal R} < 1.5$, and {\em three} free parameters in the model, $\phi_c$, $\Lambda_3$, and $\Lambda_2 \equiv V_0^{1/4}$. Then fixing the spectral index $n_{\cal R}$ and varying $\phi_c$ results in a contour in the $\Lambda_2\ /\ \Lambda_3$ plane. In addition, we have two consistency constraints: first that the dynamics of the field are dominated by the $\phi^{-p}$ term, which is equivalent to the condition
\begin{equation}
\phi < \phi_c \ll \la\phi\ra = \left[\left({p \over q + 4}\right) \Lambda_3^{p + 4}
M_{Pl}^q\right]^{1/(p + q + 4)},
\end{equation}
and second that the vacuum energy is dominated by the constant $V_0 \equiv \Lambda_2^4$, which can be expressed as 
\begin{equation}
\phi \gg \alpha^{1 / p} M_{Pl} = \left({\Lambda_3^{p + 4} \over \Lambda_2^4}\right)^{1/p}.
\end{equation}
Fig 1. shows constant $n_{\cal R}$ contours in the $\Lambda_2\ /\ \Lambda_3$ plane for an ensemble of distinct cases. 
The interesting feature of this plot is that the COBE limits on the fundamental parameters of the model are relatively insensitive to the choice of $p$ and $q$, and hence on the details of the underlying physical theory. Further, if future observations find a spectrum of density fluctuations which detectably deviates from the $n_{\cal R} = 1$ scale-invariant case, this will significantly constrain the range of allowed parameters. Of particular note is that all of the fundamental mass scales are significantly below the Planck scale, $\Lambda_2,\ \Lambda_3 \ll M_{Pl}$. Finally, it is interesting to note that the presence of nonrenormalizable terms in the Lagrangian does not significantly affect our conclusions, so that the scenario is quite robust.

\section{Conclusions}
\label{secconclusions}

A generic feature of models of nonperturbative gauge dynamics in supersymmetry is the presence of a ``scalar field'' potential of the form
\begin{equation}
V\left(\phi\right) = {\Lambda_3^{p + 4} \over \phi^p},
\end{equation}
where the field $\phi$ is in general a label for a condensate. In this paper, we consider this general form of a scalar field potential in the context of inflationary cosmology, and find that an inflationary phase in the very early universe is a generic and natural characteristic, for example, of dynamical supersymmetry breaking. We call this class of inflationary models ``Dynamical Supersymmetric Inflation (DSI).'' Like models of hybrid inflation, these models are characterized by a potential dominated by a constant term $V_0$, and require coupling to another sector to end inflation when $\phi$ reaches a critical value $\phi_c$. Unlike standard hybrid inflation models, models of this type postulate a field far from the minimum of the potential, $\phi_c \ll \la\phi\ra$. 

The primary observational constraint on models of inflation comes from the Cosmic Background Explorer (COBE) satellite, which measured fluctuations in the cosmic microwave background of magnitude $P_{\cal R} = 5 \times 10^{-5}$, with a spectral index $n_{\cal R} = 1.2 \pm 0.3$. The small fluctuation amplitude, which requires fine-tuned dimensionless parameters in typical inflation models, appears naturally here as a ratio of fundamental scales, e.g. $V_0^{1/4} \simeq 10^{10}\ {\rm GeV}$ and $\Lambda_3 \simeq 10^{6}\ {\rm GeV}$. 
We do not  pursue here the idea that the potential responsible for inflation could also arise from the 
same underlying dynamics responsible for supersymmetry breaking. It is, however, intriguing to notice that the scales $\Lambda_2$ and $\Lambda_3$ turn out to be of the right order  of magnitude to explain supersymmetric particle masses in the TeV range in the supergravity and gauge mediated supersymmetry breaking scenarios, respectively.

DSI is characterized by a ``blue'' spectral index, $n_{\cal R} > 1$. A nearly scale-invariant spectrum $n_{\cal R} \simeq 1$ is the most natural outcome, but significantly blue spectra can occur for reasonable values of the parameters. The addition of nonrenormalizable terms suppressed by powers of the Planck mass,
\begin{equation}
V = {\phi^{q + 4} \over M_{Pl}^q},
\end{equation}
does not significantly alter the inflationary properties of the potential. 

Models of cosmological inflation based on dynamical supersymmetry breaking are not only well motivated from a particle physics standpoint, but also very naturally meet constraints from observations of the CMB. These models have the unusual characteristic of possessing an {\em upper limit} on the total amount of inflation, as well as the attractive feature of predicting a ``blue'' spectrum of density fluctuations.

\section*{Acknowledgments}

This work was supported in part by DOE and NASA grant NAG5-2788 at Fermilab. We would like to thank Sandip Trivedi at Fermilab for a number of helpful discussions and for participating to the early stages of this work. 

\begin{figure}
\centerline{\epsfxsize=300pt \epsfbox{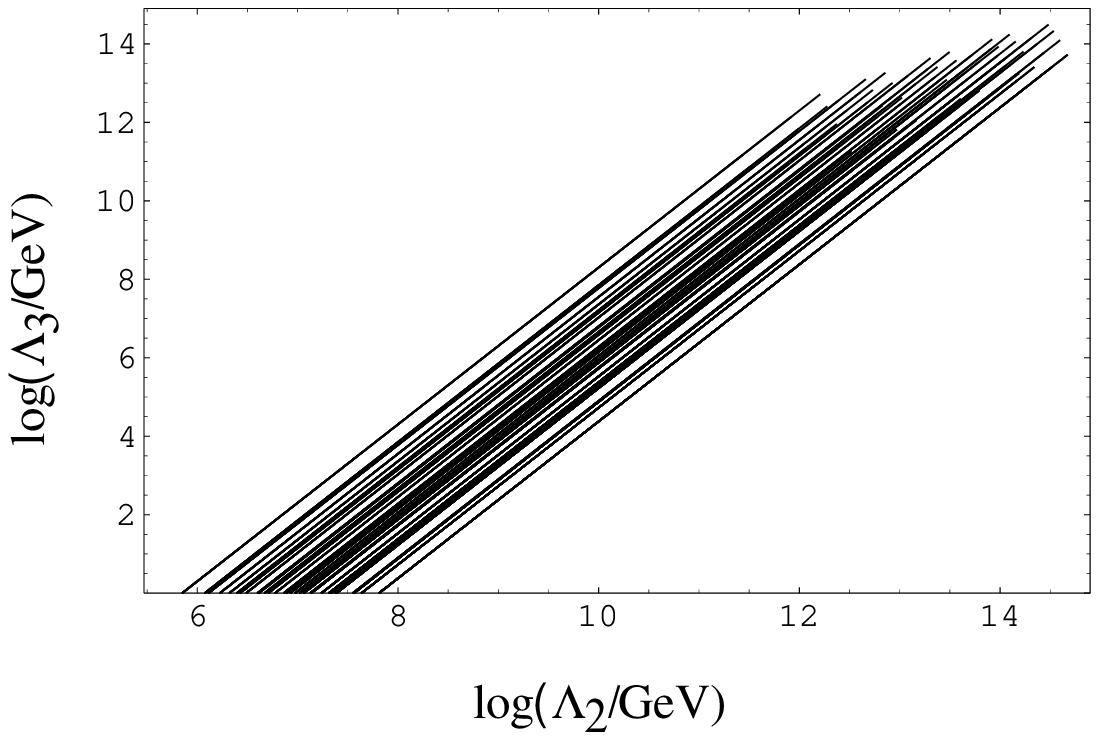}}
\hspace*{1em} {\footnotesize{{\bf Fig.\ 1:} Constant $n_R$ contours in the $\Lambda_{2} - \Lambda_{3}$ plane for two hundred cases with $p$ and $q$ varying from $2$ to $10$ and the scalar spectral index varying from $n_{\cal R} = 1.0001$ to $n_{\cal R} = 1.5$. Note that all the cases result in similar limits on $\Lambda_2$ and $\Lambda_3$, so that observational constraints are relatively insensitive to the choice of $p$ and $q$, and hence to the details of the underlying physics.}}
\end{figure}

\end{document}